\documentclass[final,3p,twocolumn,authoryear]{elsarticle}
\usepackage{lineno}
\usepackage[colorlinks=true,linkcolor=black, citecolor=blue, urlcolor=black]{hyperref}
\usepackage{amssymb}
\usepackage{natbib}
\usepackage{doi}
\usepackage{url}
\bibpunct{(}{)}{;}{a}{}{,}
\usepackage{graphicx}
\usepackage{amsmath}
\usepackage{breqn}
\def\astrobj{MW And}

\journal{New Astronomy}

\begin{document}

\begin{frontmatter}

\title{Photometric Study of Contact Binary Star MW And}

\author[ist]{Ahmed Waqas Zubairi\corref{cor1}}
\ead{ahmed.physics@gmail.com}
\author[tsu]{Shaukat Goderya}
\author[ist]{Fazeel Mahmood Khan}
\address[ist]{Institute of Space Technology, Pakistan.}
\address[tsu]{Tarleton State University, USA.}
\cortext[cor1]{Corresponding author}
%
\begin{abstract}
The Tarleton Observatory's 0.8m telescope and CCD photometer were used to obtain 1298 observations of the short period eclipsing binary star 
{\astrobj}. The observations were obtained in Johnson's BVR filters. The light curves show that {\astrobj} is an eclipsing binary star with a 
period of 0.26376886 days. Further analysis showed that the period of {\astrobj} is changing at the rate of $0.17\, sec/year$. The photometric solutions were obtained using the 2015 version of the Wilson-Devinney model. The solutions show that {\astrobj} is an eclipsing binary star of W UMa type. Our analysis suggests that the system has a light curve of W-subtype contact system. Its spectral type of K0/K1, as estimated from its color, places it in the Zero-Age contact zone of the period-spectral type diagram. Luminosity from the solutions indicates that it is a double-line spectroscopic system and therefore, spectroscopic observations are recommended for further detail study. 
\end{abstract}

\begin{keyword}
Contact Binary \sep CCD \sep Photometry
\end{keyword}

\end{frontmatter}


\section{Introduction}

W Ursae Majoris (W UMa) stars are short period eclipsing binary stars (EB) in which there is a common envelope around both stars due to overflowing Roche lobes of the stars. Inspection of All-Sky Survey data indicates that W UMa systems are very common \citep{2006Malkov}. Our understanding of their origin, structure, and evolution is vastly incomplete and as such, it has been difficult to develop a satisfactory theory for their occurrence. It is not known for sure whether the systems are born as Siamese twins or are formed from detached binaries through Angular Momentum Loss (AML) \citep{1981Vilhu}, or through Kozai  Cycle because of a third component \citep{1962Kozai}. More work needs to be done in this respect \citep{2006Paczynski}. Their lifetimes are also not very well known and various numbers from 1.0 Gyr to $>$ 5.68 Gyr are quoted in literature by different authors \citep{1992deLoore,1967Kraft}. Observational data shows that the light curve of the W UMa can be classified either as an A-type system (primary minimum due to eclipse of the larger more massive component) or a W-type system (primary minimum due to eclipse of the smaller less massive component). It has been suggested that in W-type systems, the primary component is an un-evolved main-sequence star, with later spectral type, smaller mass, lower luminosity, larger mass ratio, and a thick common envelope, while in A-type systems the primary component is approaching terminal age of the main-sequence state with earlier spectral type, larger mass, higher luminosity, smaller mass ratio, and a shallow convective envelope. Further detailed discussion on A-type and W-type can be found in various literature \citep[see for example][]{1982VanHamme,1988Hilditch,1985Rucinski}, but in short, whether they originate from the same base system, or they form from one type to the other and what will be their final fate remain unanswered questions.\\
Data on well-determined parameters of W UMa systems are few and  limited while there is no shortage of known W UMa systems from All-Sky Surveys. The problem is that for many of these systems, there is no detailed and comprehensive spectroscopic and photometric data available for the determination of absolute parameters. In addition, spectroscopic observations of faint systems, magnitude $>$ 14 require 1.5 m or larger telescopes, and considering that large telescope time is hard to secure for binary star work, it becomes much more important to first obtain photometry data to analyze the light curve and to obtain preliminary parameters of the binary system. For this reason, we have embarked on a project to obtain UBVRI photometry data on 14 and fainter magnitude W UMa systems. \\
%
In this paper, we present the photometric data analysis and modeling of {\astrobj}. The General Catalog of Variable Stars (GCVS) \citep{2017GCVS} lists {\astrobj} as a possible W UMa contact system. SIMBAD \citep{2000Wenger} search show that there are only eight references as of this writing. The available references \citep{2014Drake,2012Paschke,2006Otero} discusses the light elements of {\astrobj} only and none contain detailed photometric and modeling analysis. The references contained therein also do not provide any further information about photometric parameters. No other literature review resulted in any additional information. Therefore, we selected this system in our list of targets to observe.
\section{CCD Photometry}
\begin{figure}[!ht]
\centering
\includegraphics[width=0.65\columnwidth]{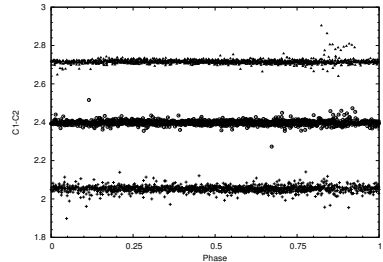} 
\caption{The difference in constant stars magnitudes is plotted with phase. The triangle is B-filter whereas the hollow circle and $+$ represents the V and R filters, respectively.}
\label{check}
\end{figure}

The finder chart of the system can easily be obtained from SIMBAD. However, we present the GSC and Tycho catalog numbers, magnitudes, and coordinates of the target and constant stars respectively in Table \ref{identification}. The constant stars are plotted with phase to check their variability in Figure \ref{check}. The Tarleton Observatory's $0.8m$ Ritchey-Chretien telescope was used to obtain $1298$ observations in 15 nights, spanning four months from November, 2012 to February, 2013. Photometry data was obtained in B, V, and R band pass filters with the help of FLI CCD camera PL4240 that has a field of view of  $0.282$ degrees in $1\times1$ binning. The exposure time of 90 second, 30 second, and 10 second in B, V, and R band pass filter, respectively, provided best fits images. Heliocentric Julian Date (HJD) and magnitude data in B, V, and R band pass was derived from the observation nights using GCX photometric reduction software \citep{2005RaduCorlan} and Python scripts provided by Goderya at Tarleton Observatory. A sample set of the observed data is shown in Table \ref{sampledata}.
\begin{table}[h]
\begin{center}
\caption{Identification data for {\astrobj}}
\resizebox{0.9\columnwidth}{!}{
\begin{tabular} {llccccc} 
\hline
Star  & Identifier & B & V & R & $\alpha(2000)$ & $\delta(2000)$    \\ 
\hline
{\astrobj} & $GSC\; 2836-1495$ & $15.00$ & $14.00$ & $13.30$ & $02\; 31\; 28.95$ & $39\; 41\; 19.3$\\ 
C1    & $TYC\; 2836-1736$ & $12.98$ & $12.49$ & $-$ & $02\; 31\; 38.95$ & $39\; 38\; 51.09$\\ 
C2    & $TYC\; 2836-1381$ & $12.44$ & $12.22$ & $-$ & $02\; 31\; 15.29$ & $39\; 39\; 02.89$\\ 
\hline
\end{tabular}
}
\label{identification}
\end{center}
\end{table}
\begin{table*}[h]
\caption{Sample of observed data for {\astrobj}} 
\centering
\begin{tabular}{cccccc} 
\hline
$MJD^{1}$ &  B & $MJD^{1}$ &  V & $MJD^{1}$ &  R\\
\hline
$6239.7765$ & $1.9525$ & $6239.7794$ & $1.6520$ & $6239.7799$ & $1.4005$\\
$6239.7807$ & $1.9880$ & $6239.7816$ & $1.6640$ & $6239.7821$ & $1.4235$\\
$.........$ & $......$ & $.........$ & $......$ & $.........$ & $......$\\
$6248.7053$ & $2.0155$ & $6248.7062$ & $1.7045$ & $6248.7067$ & $1.4475$\\
$6248.7076$ & $2.0050$ & $6248.7085$ & $1.6835$ & $6248.7090$ & $1.4465$\\
$.........$ & $......$ & $.........$ & $......$ & $.........$ & $......$\\
$6252.6667$ & $1.9760$ & $6252.6676$ & $1.6755$ & $6252.6681$ & $1.4070$\\
$6252.6740$ & $1.9415$ & $6252.6749$ & $1.6110$ & $6252.6753$ & $1.3805$\\
$.........$ & $......$ & $.........$ & $......$ & $.........$ & $......$\\
$6264.6583$ & $2.0795$ & $6264.6567$ & $1.7845$ & $6264.6572$ & $1.5220$\\
$6264.6605$ & $2.0430$ & $6264.6592$ & $1.7450$ & $6264.6597$ & $1.4680$\\
$.........$ & $......$ & $.........$ & $......$ & $.........$ & $......$\\
$6265.6446$ & $2.0070$ & $6265.6455$ & $1.7195$ & $6265.6460$ & $1.4765$\\
$6265.6478$ & $2.0605$ & $6265.6487$ & $1.7625$ & $6265.6492$ & $1.5150$\\
$.........$ & $......$ & $.........$ & $......$ & $.........$ & $......$\\
$6266.6392$ & $1.9980$ & $6266.6472$ & $1.6755$ & $6266.6477$ & $1.4220$\\
$6266.6436$ & $1.9870$ & $6266.6494$ & $1.6680$ & $6266.6499$ & $1.4030$\\
$.........$ & $......$ & $.........$ & $......$ & $.........$ & $......$\\
$6268.5569$ & $2.1825$ & $6268.5602$ & $1.9330$ & $6268.5628$ & $1.7185$\\
$6268.5593$ & $2.2140$ & $6268.5624$ & $1.9880$ & $6268.5650$ & $1.7810$\\
$.........$ & $......$ & $.........$ & $......$ & $.........$ & $......$\\
$6272.5612$ & $2.5650$ & $6272.6182$ & $1.5780$ & $6272.6209$ & $1.3625$\\
$6272.6151$ & $1.8925$ & $6272.6204$ & $1.6605$ & $6272.6231$ & $1.3675$\\
$.........$ & $......$ & $.........$ & $......$ & $.........$ & $......$\\
$6273.5486$ & $1.9600$ & $6273.5473$ & $1.6420$ & $6273.5499$ & $1.4010$\\
$6273.5508$ & $1.9865$ & $6273.5495$ & $1.6570$ & $6273.5521$ & $1.4270$\\
$.........$ & $......$ & $.........$ & $......$ & $.........$ & $......$\\
$6274.5428$ & $2.1100$ & $6274.5437$ & $1.7975$ & $6274.5465$ & $1.5125$\\
$6274.5451$ & $2.0895$ & $6274.5461$ & $1.7950$ & $6274.5487$ & $1.4845$\\
$.........$ & $......$ & $.........$ & $......$ & $.........$ & $......$\\
$6279.5646$ & $2.0100$ & $6279.5632$ & $1.7075$ & $6279.5659$ & $1.4485$\\
$6279.5668$ & $2.0180$ & $6279.5655$ & $1.7055$ & $6279.5681$ & $1.4085$\\
$.........$ & $......$ & $.........$ & $......$ & $.........$ & $......$\\
$6310.5617$ & $1.9950$ & $6310.5550$ & $1.6770$ & $6310.5630$ & $1.4360$\\
$6310.5638$ & $1.9585$ & $6310.5567$ & $1.6420$ & $6310.5652$ & $1.4370$\\
$.........$ & $......$ & $.........$ & $......$ & $.........$ & $......$\\
$6311.5375$ & $2.0290$ & $6311.5384$ & $1.7120$ & $6311.5496$ & $1.5740$\\
$6311.5439$ & $2.0985$ & $6311.5448$ & $1.8025$ & $6311.5518$ & $1.6125$\\
$.........$ & $......$ & $.........$ & $......$ & $.........$ & $......$\\
$6328.5997$ & $2.4080$ & $6328.5979$ & $2.1115$ & $6328.6060$ & $1.7115$\\
$6328.6024$ & $2.3720$ & $6328.6006$ & $2.0680$ & $6328.6082$ & $1.6695$\\
$.........$ & $......$ & $.........$ & $......$ & $.........$ & $......$\\
$6330.5907$ & $2.2125$ & $6330.5892$ & $1.9245$ & $6330.5965$ & $1.5405$\\
$6330.5929$ & $2.1745$ & $6330.5916$ & $1.8815$ & $6330.5987$ & $1.5160$\\
\hline
\end{tabular}
\label{sampledata} \\
\footnote 1 H.J.D = MJD + 2450000.0000
\end{table*} 
\begin{table*}[h]
\caption{Times of Minima for {\astrobj}. The H.J.D. are +2450000.0000}
\centering
 \begin{tabular}{cccccccccccc} 
\hline 
Calendar Date & H.J.D.      &   Wt.  &   E        &    O-C      &  Reference\\
\hline
11-12-1999 & $1523.6370$ & $7.0$ & $-17989.0$ & $-0.0014957$ & \citep{2006Otero} \\
13-08-2002 & $2500.1200$ & $3.0$ & $-14287.0$ & $ 0.0068899$ & \citep{2011O-CNelson}\\
18-12-2007 & $4452.7970$ & $5.0$ & $ -6884.0$ & $ 0.0000082$ & \citep{2012Paschke}\\
16-10-2007 & $4389.7530$ & $5.0$ & $ -7123.0$ & $-0.0031698$ & \citep{2012Paschke}\\
01-08-2016 & $7602.4610$ & $3.0$ & $  5057.0$ & $-0.0003737$ & \citep{2017Maz}\\
\textbf{B Filter}\\
21-11-2012 & $6252.7570$ & $8.0$  & $-60.0$ & $0.0000854$ & present study \\
05-12-2012 & $6266.7365$ & $9.0$  & $-7.0$  & $0.0001207$ & present study \\ 
07-12-2012 & $6268.5829$ & $10.0$ & $ 0.0$  & $0.0001178$ & present study \\ 
11-12-2012 & $6272.8030$ & $9.0$  & $23.0$  & $0.0001798$ & present study \\
12-12-2012 & $6273.5942$ & $10.0$ & $19.0$  & $0.0002094$ & present study \\ 
13-12-2012 & $6274.6495$ & $9.0$  & $23.0$  & $0.0000340$ & present study \\ 
18-12-2012 & $6279.6611$ & $8.0$  & $42.0$  & $0.0000256$ & present study \\
\textbf{V Filter}\\
21-11-2012 & $6252.7567$ & $8.0$  & $-60.0$ & $0.0000948$ & present study \\
05-12-2012 & $6266.7365$ & $9.0$  & $-7.0$  & $0.0001715$ & present study \\
07-12-2012 & $6268.5829$ & $10.0$ & $ 0.0$  & $0.0001631$ & present study \\
12-12-2012 & $6273.5942$ & $10.0$ & $19.0$  & $0.0001185$ & present study \\
13-12-2012 & $6274.6498$ & $9.0$  & $23.0$  & $0.0001156$ & present study \\
18-12-2012 & $6279.6611$ & $8.0$  & $42.0$  & $0.0000710$ & present study \\
\textbf{R Filter}\\
21-11-2012 & $6252.7568$ & $8.0$  & $-60.0$ & $0.0001558$ & present study \\
05-12-2012 & $6266.7363$ & $9.0$  & $-7.0$  & $0.0001171$ & present study \\
07-12-2012 & $6268.5829$ & $10.0$ & $ 0.0$  & $0.0001240$ & present study \\
12-12-2012 & $6273.5938$ & $10.0$ & $19.0$  & $0.0002069$ & present study \\
13-12-2012 & $6274.6498$ & $9.0$  & $23.0$  & $0.0000764$ & present study \\
18-12-2012 & $6279.6609$ & $8.0$  & $42.0$  & $0.0000956$ & present study \\
\hline \\
\end{tabular}

\label{tbl:timesofminima}
\end{table*}
\section{Period Analysis}
\begin{figure}[h]
	\centering
	\includegraphics[width=0.8\columnwidth]{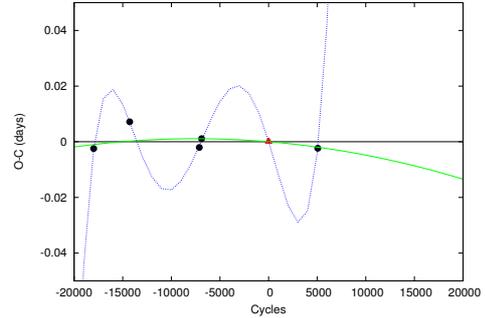}
	\caption{O-C plot in Kwee and van Woerden algorithm. The red triangle represents the observed values in V-filter. The filled circles represents the literature values and the green curve is quadratic fit for which the values are listed in Table \ref{tbl:timesofminima}. The blue dotted curve represents fifth order correction.}
	\label{O-CKewee}
\end{figure}

\begin{figure}[h]
\centering
\includegraphics[width=0.8\columnwidth]{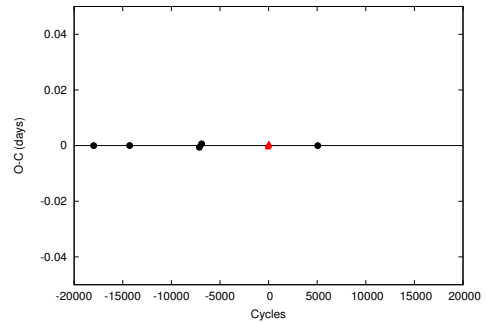}
\caption{O-C plot in Kwee and van Woerden algorithm after 5th order correction. The red triangle represents the observed values in V-filter. The filled circles represents the literature values listed in Table \ref{tbl:timesofminima}.}
\label{O-C Kewee-5}
\end{figure}

Several previously published epoch of minima can be found, these are all listed in Table \ref{tbl:timesofminima} with dates and references. The first light element was reported by \citep{2006Otero} and is shown in Equation \ref{eq:Otero}. A more precises value of the period is reported by Anton Paschke \citep{2012Paschke} and is shown in Equation \ref{eq:Paschke}. The period in Equation \ref{eq:Paschke} is also published in the Catalina Sky Survey \citep{2014Drake}. One more light element can be found from the work of Bob Nelson \citep{2011O-CNelson} and this is shown in Equation \ref{eq:Nelson}. If we compare the period in Equations \ref{eq:Otero} and \ref{eq:Paschke}, we find the difference to be $+2\times10^{-5}\, days$ (approximately $1.73\, seconds$), while the difference between the period in Equations \ref{eq:Nelson} and \ref{eq:Paschke} is $-6\times10^{-7}\, days$ (or approximately $-0.052\, seconds$).
\begin{equation}\label{eq:Otero}
Min.(I) = H.J.D.\;2451523.637\,+\,0^d.26375\:E
\end{equation}
\begin{equation}\label{eq:Paschke}
Min.(I) = H.J.D.\;2451523.637\,+\,0^d.2637700\:E
\end{equation}
\begin{equation}\label{eq:Nelson}
Min.(I) = H.J.D.\;2452500.1120\,+\,0^d.2637694\:E
\end{equation}
The observations reported in this paper contains $19$ more epoch of minima in B, V and R wavelengths. These are also shown in Table \ref{tbl:timesofminima}. The epoch of minima were obtained using the the Minima25C software. This software is publically available and is authored by Bob Nelson \citep{2005Nelson}. All algorithms provided in this software were used to derive the epoch of minima, however, only the Kwee and van Woerden algorithm \citep{1956Kewee} provided the best epoch data. With all the epoch of minima listed in Table \ref{tbl:timesofminima}, it is possible to perform determination of new light elements and see if there is any period change in the system. Of our $19$ epoch of minima, only the ones in V band pass were used as the cycles of other bandpass are similar and do not change the outcome of the computations. We used the method of generalized least square. Equation \ref{eq:firstorder} shows the first order linear least square calculations. The O-C (days) vs. E (cycles) plot is shown in Figure \ref{O-CKewee}. Inspection of Figure \ref{O-CKewee} shows that there is considerable variation in the O-C values, therefore a second order corrections was attempted. The light elements are shown in Equation \ref{eq:secondorder} and the computed O-C curve is shown in Figure \ref{O-C Kewee-5} as quadratic curve. While the O-C values derive after second order correction do not fit well, none the less the coefficient of the $E^2$ term allows us to calculate $\Delta P/P$.  Higher order corrections were also attempted to obtain more accurate representation of O-C values. Figure \ref{O-C Kewee-5} also shows the 5th order corrections and Equations \ref{eq:fifthorder} shows the light elements. If we compare the period in Equation \ref{eq:fifthorder} with Equation \ref{eq:Nelson}, the difference is $-1.2\times10^{-5}\, days$ (approximately $-1.01$ seconds). Analysis show that the period is decreasing at the rate of $0.17\, sec/year$. At present, the number of epoch data points is too few to justify accepting 5th order correction, therefore lights elements from second order correction are used for constructing the light curve for modeling purpose. In addition, the effect on overall light curve is extremely small and thus would not affect the modeling process.
%

\begin{dmath}\label{eq:firstorder}
		Min.(I) =  H.J.D.\;2456268.5827\,(\pm 7) 
		 +\,0^d.26376915\,(\pm 2)\:E 
\end{dmath}

\begin{dmath}\label{eq:secondorder}
 Min.(I) =  H.J.D.\;2456268.5829\,(\pm 7)\\
  +\,0^d.26376886\,(\pm 2)\:E 
  -\,1.9\times10^{-11}\:E^{2} 
  \end{dmath}
\begin{dmath}\label{eq:fifthorder}
 Min.(I) =  H.J.D.\;2456268.5829\,(\pm 7) \\  +\,0^d.26375772\,(\pm 2)\:E -\,9.03\times10^{-11}\:E^{2} \\  +\,3.49\times10^{-13}\:E^{3}
 +\,4.41\times10^{-17}\:E^{4} \\ +\,1.32\times10^{-21}\:E^{5}
   \end{dmath}

\section{Photometric Analysis}
\begin{figure}[h]
	\centering
	\includegraphics[width=0.8\columnwidth]{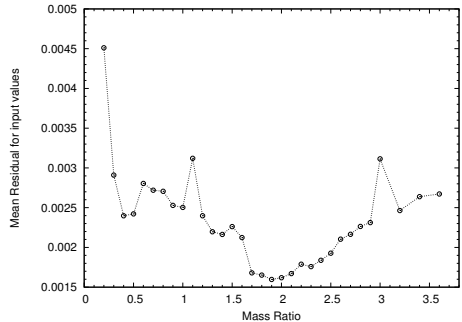}
	\caption{Mean residuals versus the mass ratio for {\astrobj}. Each circle represents a solution.}
	\label{gmr}
\end{figure}

The $2015$ version of the Wilson-Devinney program (WD) \citep{1971Wilson-Divinney,2014Wilson-VanHamme} was used in light curve modeling. Approximation of a function using the Fourier technique was used to generate the normalized light curve to speed up the modeling process \citep{1993Rucinski}. The normalized light curve was superimposed on observed data to visually verify the validity of the normalized light curve. The weight of each night was based on the parabolic shape of the minimum. Inspection of the light curve indicates that the system is either mode $1$ or $3$ of the Wilson-Devinney model; however, trial solutions for the semi-detached mode were also performed but only mode $3$ provides the best visual fit to the observed light curve. Therefore, mode $3$ was adopted for further modeling with eccentricity $e=0$ and rotation parameters ($F_1$, $F_2$) equal to $1$. Literature review for spectral classification was not positive; thus, the only way to estimate the temperature of the hotter component is the B-V index \citep{2003royastronomy} assuming no reddening. The modeling process started with the light curve fitting of the normalized observed data with the LC unit of the WD model to obtain the initial estimates of the photometric parameters. This was done for each bandpass filters B, V, and R. The initial estimates from LC were then used to produce input files for the Differential Correction (DC) unit of the WD model for simultaneous modeling. \\
The DC unit was first used to search for the global mass ratio. This is typically known as ``q search procedure'' in WD modeling. In this search, $32$ solutions were obtained for discrete values of the mass ratio. The results are displayed in Figure \ref{gmr} and it shows that the global mass ratio is  $q=1.90$. After the q search procedure, modeling with the DC unit started with keeping the coarse parameters ($i$, $\Omega_1$, $q(m_2/m_1)$, $T_2$, $L_{B,1}$, $L_{V,1}$, $L_{R,1}$) adjustable. Upon obtaining a converging solution with the smallest mean residuals, the next step was to allow adjustment of minor parameters ($x_1$, $x_2$, $A_1$, $A_2$, $g_1$, $g_2$) by keeping the third light ($l_3$) zero. After many trials, limb darkening ($x_1$,$x_2$), gravity darkening ($g_1$, $g_2$) parameters converged with an improved mean residual for the input values. All of the other free coarse parameters also converged properly. Table \ref{parameters} shows the adopted photometric parameters for {\astrobj} without any spot; henceforth, it is referred to as a no-spot model.\\
Superimposing the no-spot model on the observed light curve showed that the fit on the secondary maximum is not good. This indicated that there might be star spots present on either of the stars and therefore spot modeling was performed. The spot model solutions consist of two steps. As a first step, the adopted photometric parameters with the no-spot model from Table \ref{parameters} serve as the input parameters for the LC unit to find the coarse values of the spot parameters. In the trials that were performed, two spots were chosen: one on star 1 and another on star 2. In the second step, the spot parameters from the LC unit were inserted into the input file for the DC unit. The solution trials began by keeping the coarse, minor parameter as mentioned above free and adjustable including the spot parameters latitude, longitude, angular radius, and temperature factor. Again after many trials, convergence is found only for the temperature factor and longitude of the spots. Table \ref{parameters} shows the list of all parameters found with minimum mean residual while Table \ref{spotparameters} shows the spot parameters adjusted by DC. Figure \ref{configuration} clearly shows one spot on each component of the {\astrobj}.
\section{Light Curve Analysis}
\begin{figure}[h]
\centering
\includegraphics[width=\columnwidth, height=9cm]{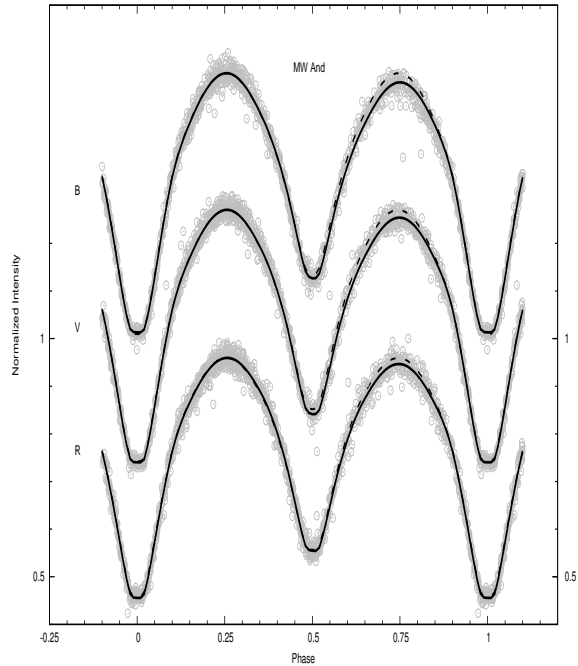} 
\caption{The figure shows observed and computed light curves for both solutions of {\astrobj}. The circles are individual observations while the dash-line is the no-spot solution. The continuous black curve is the spot solution.}
\label{lcandcheck}
\end{figure}

\begin{figure}[h]
\begin{center}
 \includegraphics[width=0.30\columnwidth, height=2.90cm]{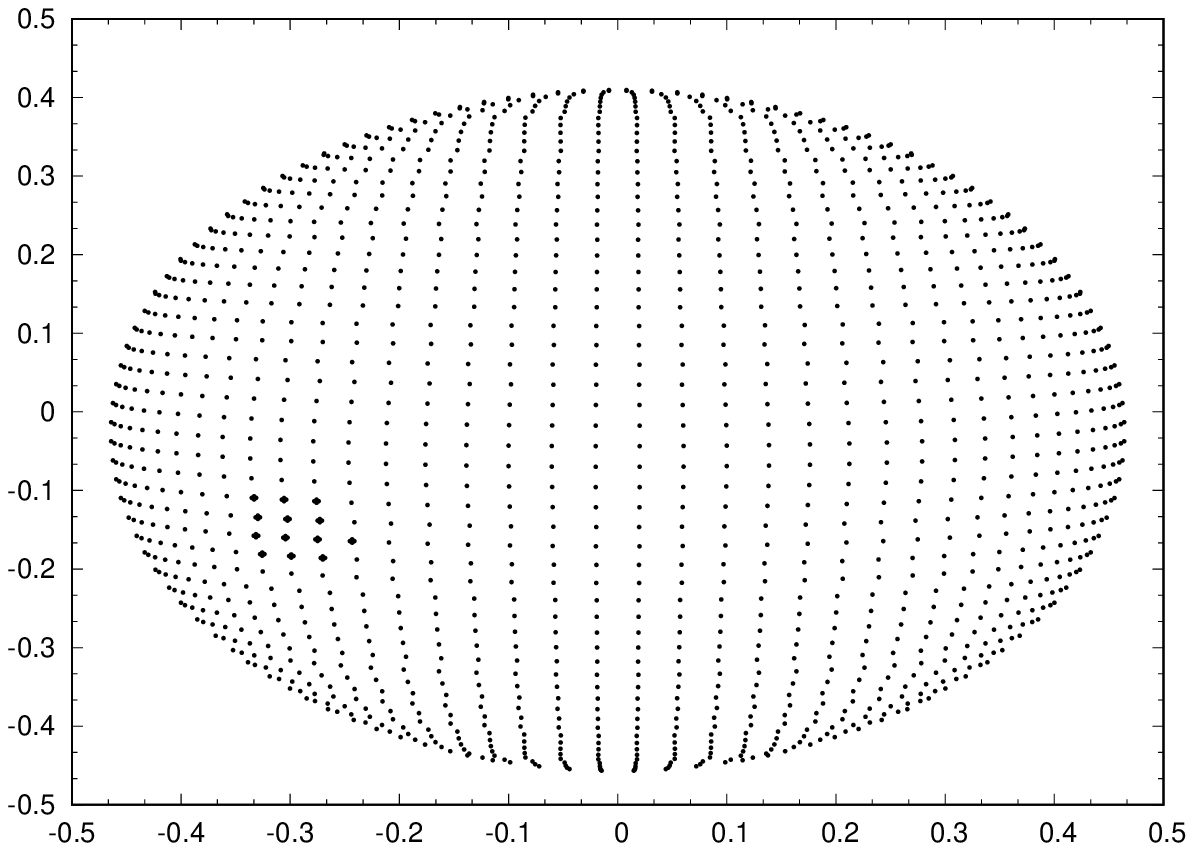}
 \includegraphics[width=0.60\columnwidth, height=2.90cm]{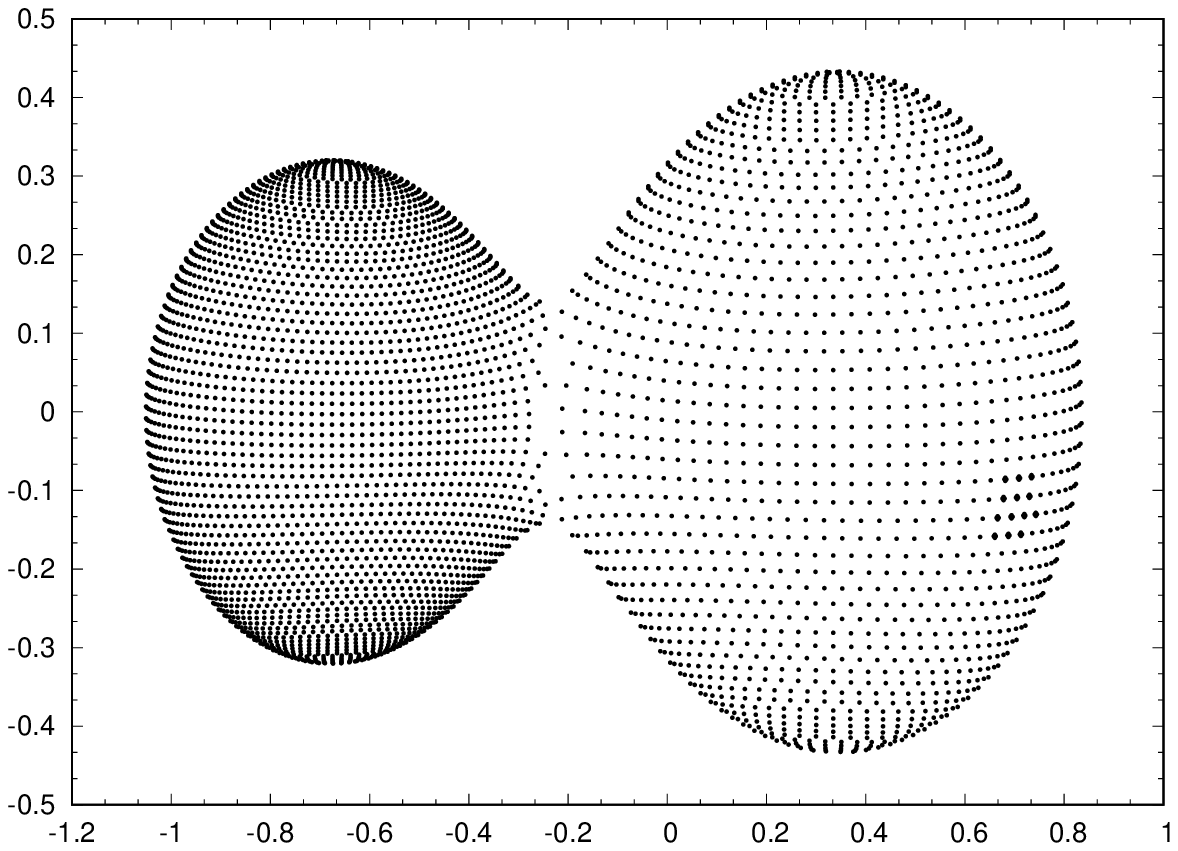} \\
 \includegraphics[width=0.30\columnwidth, height=2.90cm]{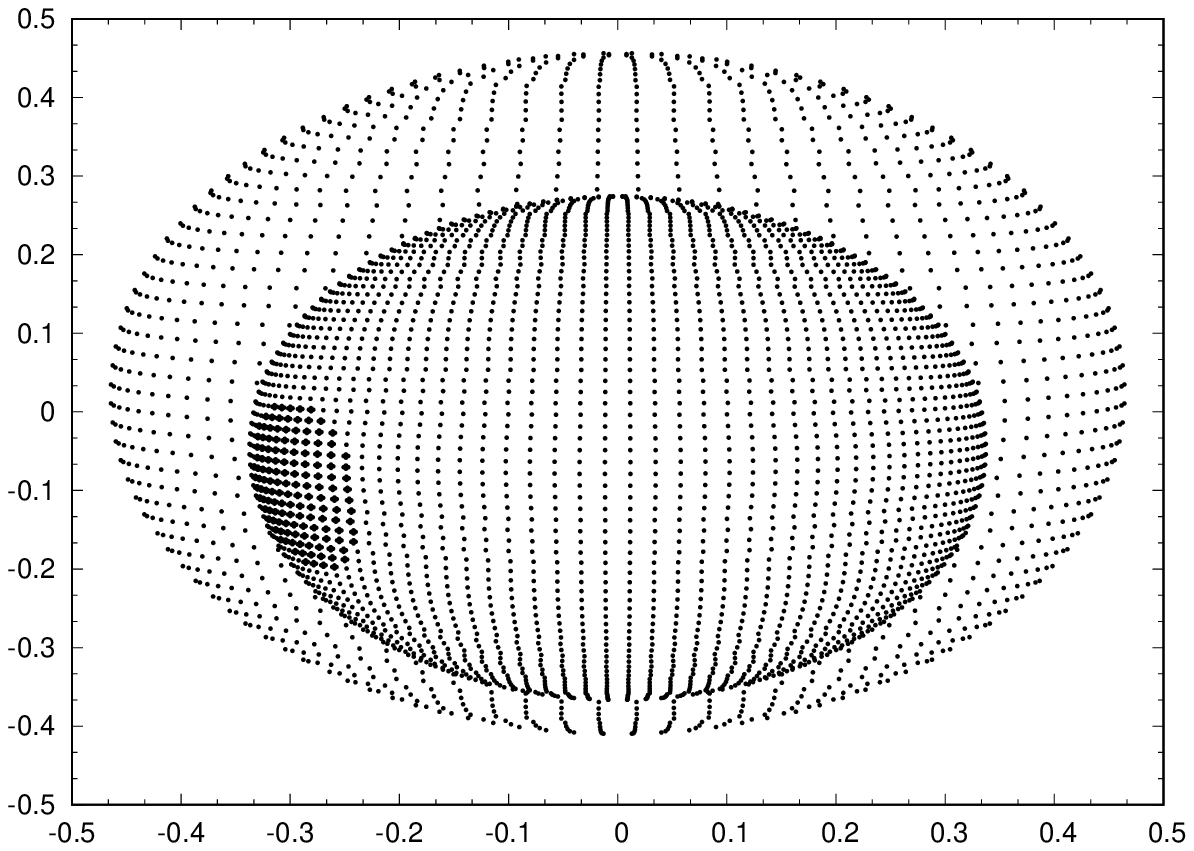}  
 \includegraphics[width=0.60\columnwidth, height=2.90cm]{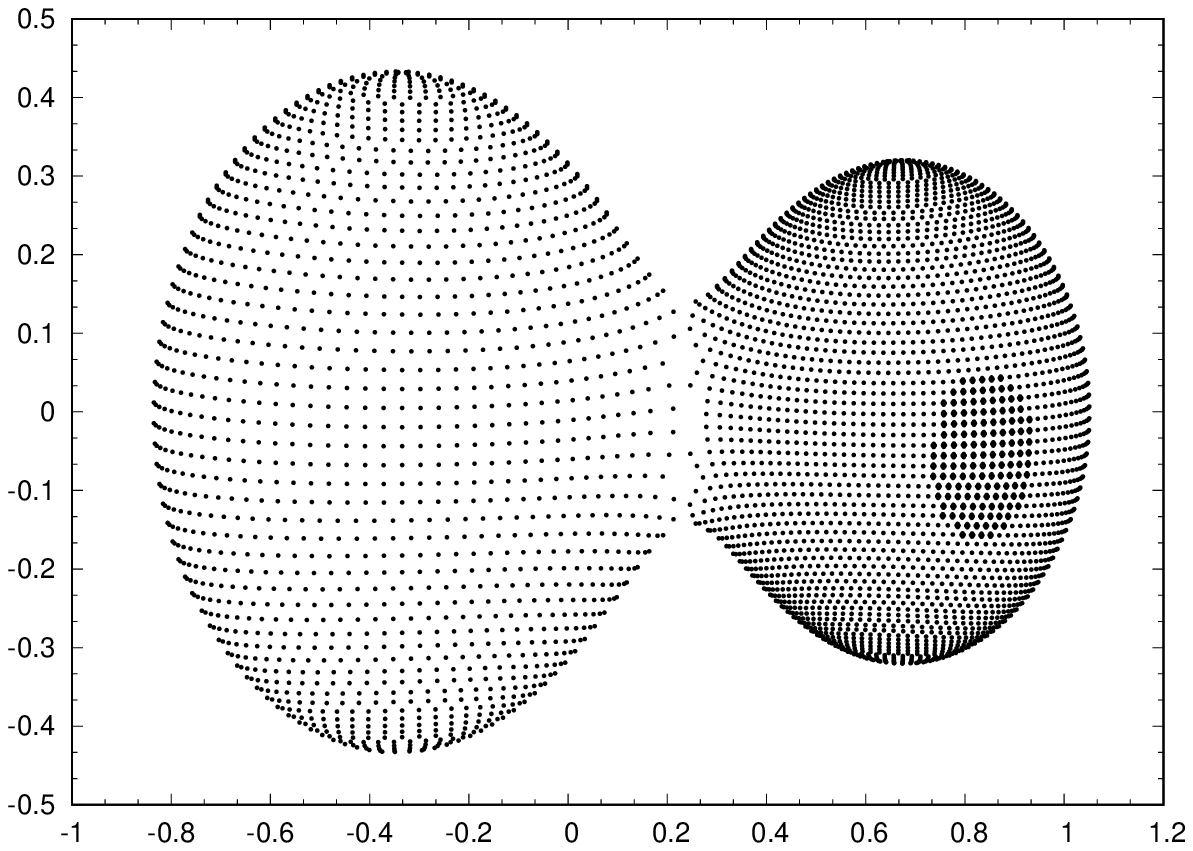} 
\caption{The top panel shows configurations of {\astrobj} at phase $0$ and $0.25$ while the lower panel shows $0.50$ and $0.75$ respectively.} 
\label{configuration}
\end{center}
\end{figure}

Figure \ref{lcandcheck} shows the observed light curves B, V, and R colors together with the computed curves for the spot and no-spot models. The dash-line shows the no-spot solutions whereas the solid line represents the spot solutions. Figure \ref{lcandcheck} shows that both solutions fit the observed data between phases of 0.0 to 0.5 whereas between 0.5 to 1.0 phase, the spot solution fits better. It appears that there is a small flat portion in the light-curve at primary minimum which may be an indicative of total eclipse. Analysis shows that {\astrobj} is a W UMa-type contact binary star with a mass ratio of $q = 1.99$ (or $1.97$ for no-spot). The temperatures of two components differ by at least $220$K for the no-spot solutions and $241$K for the spot solution. The temperature of the secondary is smaller than the primary whereas its luminosity is larger than the primary. The configuration in Figure \ref{configuration} show that the secondary is also larger in size compared to the primary. This indicates that {\astrobj} is a W-type contact binary system.  {The percentage of contact is between $40.15\%$ to $44.99\%$ in the two solutions.} The two maxima show a small difference in luminosity and this could be an  indicative of the poorly studied O'Connell effect \citep{1951Connell,2009Wilsey}. From the luminosities in Table \ref{parameters}, it appears that {\astrobj} may be a double-line spectroscopic system. 
\begin{table*}[h]
	\begin{center}
	  \caption{Photometric solution for {\astrobj}}
\begin{tabular}{ p{4cm} p{3.5cm} p{3.5cm} }
	\hline
	Parameters & No Spot & Spot \\
	 \hline

$q (m_2/m_1)$ & 1.97 $\pm$ 0.01 & 1.99 $\pm$ 0.01\\ 
i (deg)      & 86.12 $\pm$ 0.44 & 86.10 $\pm$ 0.22 \\ 
$\Omega_1 = \Omega_2$   & 4.983 $\pm$ 0.019 & 5.011 $\pm$ 0.008 \\
$\Omega_{in}^{2}$  & 5.251689 & 5.251689 \\
$\Omega_{out}^{2}$ & 4.654520 & 4.654520 \\
$f (\% \, of\, overflow)$  & $44.99\%$ & $40.15\%$ \\
$L_1/(L_1+L_2)(B)$ & 0.443 $\pm$ 0.004 & 0.450 $\pm$ 0.001 \\
$L_1/(L_1+L_2)(V)$ & 0.431 $\pm$ 0.003 & 0.436 $\pm$ 0.001 \\
$L_1/(L_1+L_2)(R)$ & 0.420 $\pm$ 0.002 & 0.423 $\pm$ 0.001 \\
$T_1^{4}(K)$    & 4579 & 4579 \\
$T_2(K)$    & 4359 $\pm$ 11 & 4338 $\pm$ 4\\
$A_1^{3}$    & 1.00 & 1.00 \\
$A_2^{3}$    & 1.00 & 1.00 \\
$l_{3B,3V,3R}$     & 0    &   0 \\
$x_1(B)$    & 0.522 $\pm$ 0.054  & 0.565 $\pm$ 0.024\\
$x_1(V)$    & 0.340 $\pm$ 0.051  & 0.388 $\pm$ 0.022\\
$x_1(R)$    & 0.121 $\pm$ 0.048  & 0.172 $\pm$ 0.022\\
$x_2(B)$    & 0.341 $\pm$ 0.046  & 0.364 $\pm$ 0.021\\
$x_2(V)$    & 0.226 $\pm$ 0.042  & 0.251 $\pm$ 0.020\\
$x_2(R)$    & 0.065 $\pm$ 0.039  & 0.086 $\pm$ 0.021\\
$g_1$       & 0.089 $\pm$ 0.031  & 0.135 $\pm$ 0.010\\
$g_2$       & 0.239 $\pm$ 0.033  & 0.189 $\pm$ 0.014\\
$r_1(pole)$ & 0.3209 $\pm$ 0.0010 & 0.3206 $\pm$ 0.0004 \\
$r_1(side)$ & 0.3382 $\pm$ 0.0012 & 0.3381 $\pm$ 0.0005 \\
$r_1(back)$ & 0.3861 $\pm$ 0.0020 & 0.3870 $\pm$ 0.0009 \\
$r_2(pole)$ & 0.4320 $\pm$ 0.0022 & 0.4340 $\pm$ 0.0009 \\
$r_2(side)$ & 0.4634 $\pm$ 0.0031 & 0.4659 $\pm$ 0.0013 \\
$r_2(back)$ & 0.5002 $\pm$ 0.0046 & 0.5030 $\pm$ 0.0020 \\ \\

Mean Residual & 0.001152 & 0.000585 \\

\hline
\footnote 2Theoretical Values & \footnote 3Assumed
\end{tabular}
\label{parameters}
\end{center}
\end{table*}

\begin{table*}[h]
	\begin{center}
		\caption{Star spot parameters for {\astrobj}}
	\begin{tabular}{lcc} 
			\hline
			  Parameters & Star Spot 1 & Star Spot 2   \\ 
			\hline
			Latitude$^{4}$(rad)         & 1.30 & 1.77\\ 
			Longitude     (rad)         & 1.68 $\pm$ 0.02 & 2.47$^{4}$ \\
			Angular Radius$^{4}$ (rad)  & 0.295 & 0.096 \\
			Spot Temp. Factor           & 0.894 $\pm$ 0.002 & $0.800^{4}$ \\
			\hline
\footnote 4Assumed
		\end{tabular}
		\label{spotparameters}
	\end{center}
\end{table*}

\section{Discussion and Conclusion}
\citet{1985Maceroni} from their study of 42 W UMa system have shown W UMa stars evolve from high to low mass ratio. A-type and W-type systems are believed to be in slightly different states of evolution. \citet{1988Hilditch} proposed that W-type systems evolve into A-type systems, while \citet{2006Gazeas} proposed the opposite. {\astrobj} show W-type light curve at present, with evolve components. With the current data, it is not possible to predict its future evolution.\\
Period changes in contact binaries is attributed to three different causes: 1) mass exchange and/or mass loss, 2) apsidal motion, and 3) the possibility of a third body. Inspection of Table \ref{parameters} show no presence of a third body or apsidal motion, therefore the period decrease in the system is most likely due to mass exchange and/or mass loss. The intense magnetic field as evident by presence of star spots control the mass flow and magnetic breaking \citep{1988Guinan}. {\astrobj} contains star spot on each star so this further supports mass exchange or loss. \\
According to the popular view of the stellar evolution and structure of contact binary stars contact system evolve from detached systems to semi-detached and then finally to contact state \citep{1992deLoore, 1993RealmofInteractingBinary, 2018newmechanismforWUMa, 1996goderyaV719Her}. Mass transfer can occur during the core H-burning phase (Case A type mass transfer) or during the shell H-burning phase (Case B type mass transfer). Detailed discussion on mass transfer can be found elsewhere \citep{1967Kippenhahn, 1968Palvec, 1971Paczynski}. Most contact systems are thought to be main sequence stars however, considering the spectral type of {\astrobj} and its position in the period-spectral type diagram for contact binaries \citep{1995goderyaV508Cyg, 1975Yamasaki}, it appears that {\astrobj} is a Zero-Age contact system with case A-type mass transfer. \\
From our study of {\astrobj}, we conclude that it is a high mass ratio contact binary system in zero age contact phase and currently going through a period change of 0.17 sec/year. The luminosity of each component is about the same, indicating the possibility of double-lined spectroscopic system. We therefore propose that radial velocity data for this system should be obtained to accurately determine the temperature from its spectral class and compute the absolute dimensions from the combined analysis of photometric and radial velocity data. Very few zero age contact binary systems have been discovered so far and for this reason it becomes a very interesting candidate to choose for future studies.
\section*{Acknowledgments}
This research has made use of the SIMBAD database, operated at CDS, Strasbourg, France, the observatory, and the computing facilities of Tarleton State University, Stephenville, TX, USA, and the office facilities of the Institute of Space Technologies, Islamabad, Pakistan. The authors would like to thank Dr. Wilson for releasing the WD code in the public domain. Thanks and credits are due to the creators of the following software: SIMBAD, GCX, and Minima25C. Finally, the authors would like to thank the anonymous reviewers for giving valuable suggestions and comments that resulted in considerable improvement of the manuscript.
\section*{References}
\bibliography{paperref-short} 
\end{document}